# Monte Carlo calculations for simulating electron scattering in gas phase

Lukas Pielsticker[1], Robert Schlögl,[1] Mark Greiner[1]

[1] Max-Planck Institute of Chemical Energy Conversion, Stiftstrasse 34-36, 45470, Mülheim an der Ruhr, Germany

**0. Abstract**

Here we present the derivation, description and results of a Monte Carlo-based algorithm for simulating inelastic scattering of photo-electrons when passing through some scattering medium, such as a gas atmosphere or a solid material. The code used to run these simulations was written in python and is freely available online. [1]

**1. Introduction**

The measurement method X-ray photoemission spectroscopy (XPS) is often used to determine the chemical state (such as oxidation state and chemical bonding environment) of solid surfaces. During a measurement, ionizing X-ray radiation is projected onto a sample, causing the electrons to be emitted from the sample's surface. The emitted electrons have a distribution of kinetic energies. By analyzing the kinetic energy distribution–using devices like hemispherical analyzers–one can determine the chemical composition and oxidation state of the sample's surface atoms. Due to the need to preserve the kinetic energy distribution of the emitted electrons, such measurements are typically performed with the sample housed in a vacuum environment (pressure ca. 5E-9 mbar); otherwise atmospheric gas molecules would inelastically scatter the emitted electrons, changing their kinetic energy distribution.

A modern variant of XPS, commonly called near-ambient pressure XPS (NAP-XPS) can measure samples in the presence of a gas atmospheres at pressures around 1-100 mbar, using a configuration that minimizes the distance electrons must travel through the gas atmosphere. [2] Nonetheless, the presence of the gas causes inelastic scattering of electrons, resulting in lower signal intensity and 'ghost' signals in the spectrum. [3] A population of electrons, having some kinetic energy distribution, will have a new kinetic energy distribution after inelastically colliding with gas molecules. If every electron in the population inelastically collides one time during the interaction, then the kinetic energy distribution after the collision, will be the convolution of the initial kinetic energy distribution, with the so-called energy-loss function of the gas molecule, as illustrated in Figure 1. The energy-loss function represents the probability distribution of all possible amounts of energy loss for the collision event (in figure 1C, the energy loss of zero is not shown).

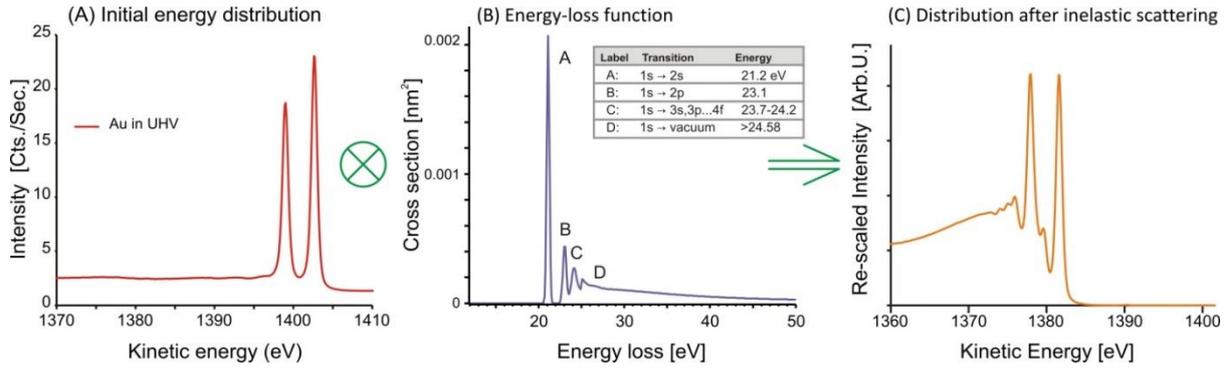

*Figure 1 - An illustration of how the kinetic energy distribution of electrons after exhibiting inelastic scattering, is the result of convolving an primary kinetic energy distribution (A) with an energy-loss function (B). The distribution shown in (C) would be the result of all the electrons from the distribution in (A) were inelastically scattered exactly once, where the amount of energy lost is chosen from the distribution in (B).*

If a population has been scattered exactly *n* times, then the resulting energy distribution will be the initial kinetic energy distribution, convolved with the energy-loss function *n* times. These line-shapes result in 'ghost lines' in the measured spectrum. In order to remove the 'ghost lines', we must know the probability of an electron being scattered *n* times after passing through a finite distance of gas phase. This probability distribution can be modeled using Poisson statistics:

$$P_n = \frac{1}{n!}\left(\frac{t}{\lambda}\right)^n e^{-t/\lambda} \qquad \text{equation 1}$$

where $P_n$ is the probability of an electron being scattered *n* times after passing through some distance *t* of scattering medium, where the inelastic mean free path is *λ*. The problem with using the Poisson equation to model inelastic scattering is that the term λ is actually a function of electron kinetic energy, and after each energy-loss event, the kinetic energy changes. Furthermore, in the experimental measurement configuration, *t* (i.e. path length) is not a single value, but rather a distribution, as illustrated in Figure 2.

The purpose of the present work is to model the experimental situation by simulating electron scattering, where the path length distributions and behavior of λ as a function of kinetic energy are included, and compare the results with those obtained by using the simple Poisson equation (with a constant λ and scalar value for t). For the simulation we use Monte Carlo methods, where changes in direction on scattering, energy losses, and changes to λ on energy loss, as well as elastic scattering events, are all included in the model.

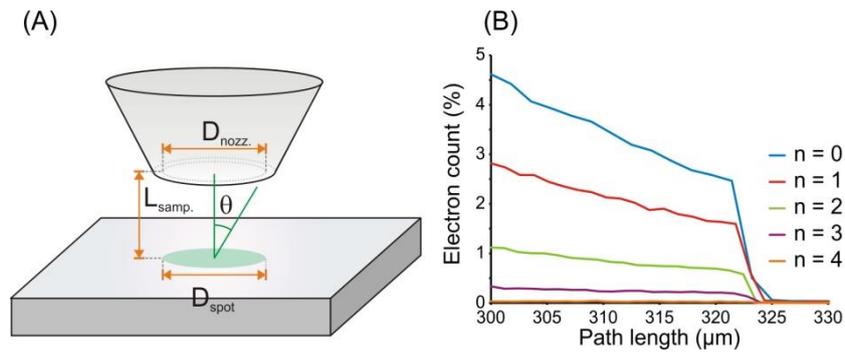

*Figure 2 (A) Schematic representation of the sample-nozzle configuration for a NAP-XPS measurement. The indicated quantities are sample-nozzle distance ($L_{samp.}$), nozzle diameter ($D_{nozz.}$), X-ray spot diameter ($D_{spot.}$) and acceptance angle ($\vartheta$). (B) A plot of electron path lengths resulting from a Monte Carlo simulation of electrons (having initial kinetic energy of 1100 eV) travelled through 25 mbar of He, where the nozzle has an acceptance angle of 22°, the sample-nozzle distance is 0.3 mm, the nozzle diameter is 0.3 mm, and the X-ray spot size is 0.3 mm.*

The plots above show (A) a schematic representation of the sample, nozzle and illuminated spot (i.e. source of electrons), (B) the path length distributions of electrons from a simulation of 2 million electrons, having initial kinetic energy of 1100 eV, passing through 25 mbar of He, with a sample-nozzle distance ($L_{samp}$) of 0.3 mm, nozzle diameter ($D_{nozz}$) of 0.3 mm and diameter of the illumination spot ($D_{spot}$) of 0.3 mm, and an acceptance angle (θ) of 22°. From this simulation it is clear that the path length travelled from the sample to the nozzle is not a single value, but rather is a distribution of values ranging from 0.30 to 0.325 mm.

## 2.1 Description of the model

The following text describes the model used to simulate photoelectron scattering. The model has been implemented in an algorithm, and used to determine scattering statistics (such as number of electrons that have been inelastically scattered *n* times along their paths), path length distributions, and angular distributions for simulations of photoelectrons traveling through a scattering medium. The methods have been applied to gas-phase scattering media, to simulate the situation encountered during near-ambient pressure XPS measurements; however, they can also be applied to modeling electron scattering in solids, and examples are available online. [1] Here we provide the derivation of all equations and methods used in the algorithm. The algorithm was written in Python 3.8 and is available online at Github. [1]

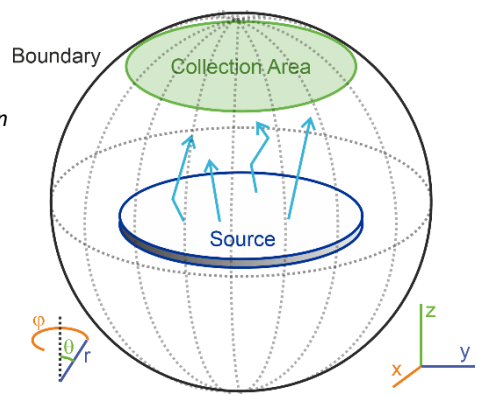

*Figure 3 - Schematic of the configuration used for Monte Carlo simulations. Electrons are generated at the source. They travel through the scattering medium, until they intersect with the boundary. To simulate a measurement, electrons can be selected from those who have intersected the boundary within the collection area.*

The simulation takes place in a geometric environment, as illustrated in Figure 3, consisting of a sphere, which we will call our *boundary*, and a disc centered inside the boundary, which we will call our *electron source*. In the case where we are simulating electron scattering in a near-ambient pressure XPS measurement, the volume enclosed by the boundary would be considered filled with a gas phase *scattering medium*. The scattering medium has the attributes *density*, *energy-loss function*, *inelastic scattering cross section* and *elastic scattering cross section* (which are functions of the moving electron's kinetic energy). In the model, electrons are generated at the source, travel until they hit the boundary. Electrons that intersect the boundary inside the *collection area* are considered to be measured electrons, and are subsequently counted and analyzed.

The general steps involved in the algorithm for simulating scattering through gas phase are as follows:

1. Randomly choose a position on the upper surface of the source as the point of creation of an electron. We give the electron a starting kinetic energy, chosen from a Gaussian-distribution of energies.
2. Choose a random direction for the electron by selecting randomly from a Lambertian angle distribution. The Lambertian is chosen because photoemission is known to follow Lambertian emission. Alternatively, one could assume non-Lambertian, and first simulate electron generation and scattering inside the solid, however, this is computationally much more expensive, and from our tests, we have found it to indeed give rise to a Lambertian profile).
3. Choose a random distance from an exponential decay distribution (see section 5.2 for further explanation). The exponential distribution is determined by the scatterer's density and total scattering cross section. The chosen distance represents how far the electron travels before it is scattered.
4. Update the electron's position by moving along its velocity vector by an amount equal to the randomly chosen distance.
5. Randomly choose whether the electron is elastically or inelastically scattered. The probabilities are based on the relative elastic and inelastic scattering cross sections, which are functions of the electron's kinetic energy, and are determined by fits to reference data. [4]
6. If the electron is inelastically scattered, choose randomly the amount of kinetic energy lost, by selecting from a distribution that resembles the scatterer's energy-loss function. Loss functions for many solids can be found in available literature. [5]
7. Choose a new direction of motion by randomly selecting from an angular spread distribution, based on the Rutherford scattering model. The form of the angular spread distribution depends on whether the electron is inelastically or elastically scattering.
8. Update the electron's velocity vector, based on its updated angle and updated kinetic energy.
9. Repeat steps 3-8 until the electron's position is outside of the boundary.
10. Record the position where the electron intersects the boundary.
11. Repeat steps 1-10 for the desired number of electrons.
12. Count all the electrons that intersect the sphere in the region bound by the *collection area*.

Note: the algorithm used here was not optimized for speed, but rather optimized for conceptual clarity and flexibility. For more computationally efficient algorithms see Werner. [6]

## 2.2 Fitting scattering cross-section functions

In order to update the total elastic scattering and total inelastic scattering cross sections every time the electron kinetic energy changes, we need to have a function that relates kinetic energy to cross section. It is known that a power law can fit the behavior well for energies > 100 eV and energies < 2000 eV. [7] Therefore, we have used published data of cross sections at various kinetic energies and fit them to power law functions. For elastic scattering, we have simply taken values from the NIST electron scattering database. [4] The data is provided in units of Bohr radii. We have converted the values to nm². We fit the data with a curve of the form:

$$\sigma = mE^{-k}$$

equation 2

where $\sigma$ is the cross section in nm², $m$ is a pre-factor, $E$ is the kinetic energy of the electrons in eV, and $k$ is an exponent factor. For the case of He as the scatterer, we used the values $m$ = 0.95 and $k$ = 1.12

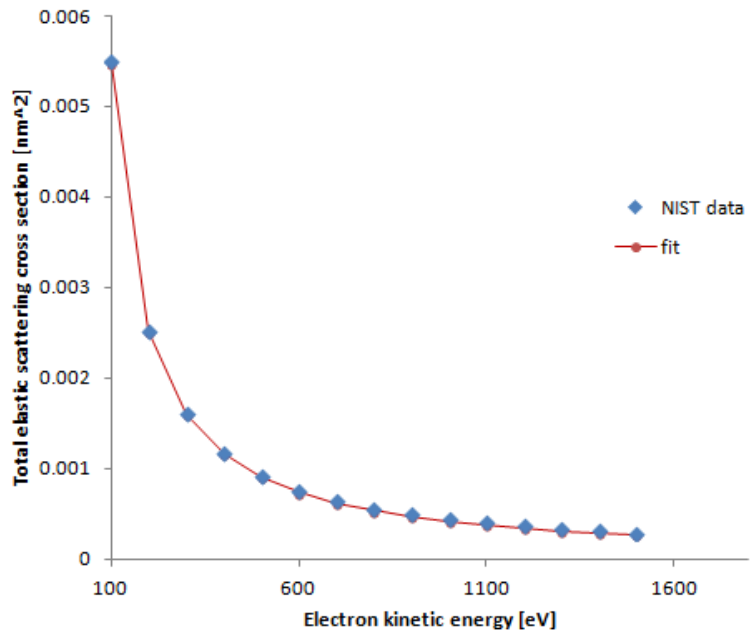

*Figure 4 - Plot of total elastic scattering cross section versus electron kinetic energy based on data from the NIST database, plotted with a curve fit to the data using equation 2.*

For the total inelastic scattering cross section, we took values from the QUASES IMFP calculator, [8] which reports inelastic mean free paths for many materials, based on the TPPM2 algorithm. [9] In order to convert the calculated mean free paths into cross sections, we needed to first convert from Å to nm, then multiply the IMFP by the density, which for the QUASES software, was the density in g/cm³ of the liquid phase. Then we take 1/(IMFP * density) as the cross section.

The cross section was again fit using a curve of the form $\sigma = mE^{-k}$. For Helium, we used values of $m$ = 0.44 and $k$ = 0.83.

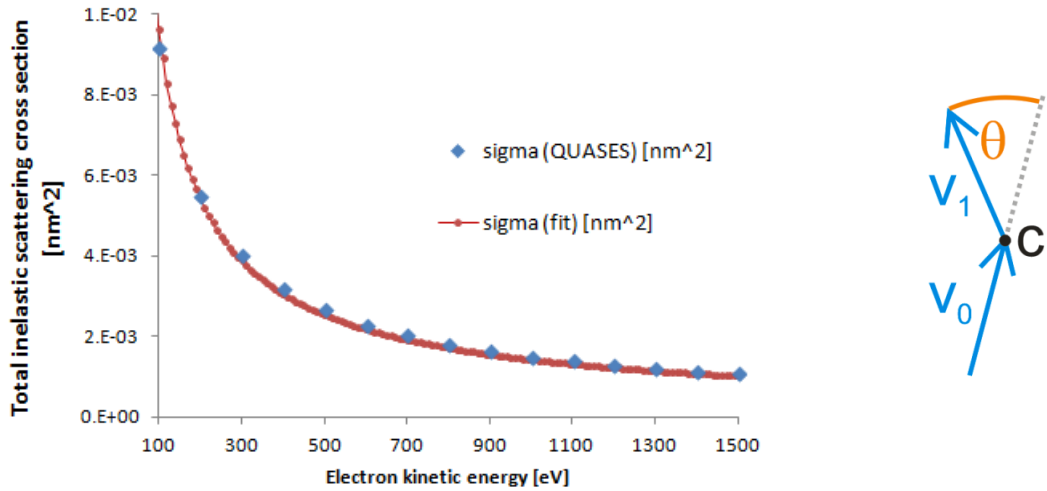

*Figure 5 - (left) Plot of total inelastic scattering cross section, as calculated using the QUASES software (using the TPPM2 algorithm), plotted with a fit to the data using equation 2. (right) a schematic illustrating electron deflection during scattering, where $v_0$ is the initial electron direction, $v_1$ is the direction after scattering, C is the point of collision, and $\vartheta$ is the polar angle of deflection.*

A simple analytic function was also needed for the differential elastic scattering cross section. The differential elastic scattering cross section represents an angular distribution function for the polar angle into which an electron is deflected after an elastic collision. The electron, travelling with initial velocity $v_0$ is defected to velocity $v_1$ after colliding with a particle at the point $c$. [10] The polar angle $\theta$ and azimuthal angle $\phi$ of deflection are relative to the electrons initial velocity. The distribution of azimuthal angles is uniform, while the distribution of polar angles is not. A function to describe the distribution of polar angles was derived by Rutherford. [11]

$$\frac{d\sigma}{d\Omega} = \frac{Z^2 e^4}{4E^2(1 - \cos\theta + 2\beta_N)^2} \qquad \text{equation 3}$$

Where Z is the atomic number, $e$ is the electron charge, $E$ is the electron kinetic energy, and $\beta$ is the atomic screening parameter, given by :

$$\beta_N = \frac{5.43 Z^{2/3}}{E} \qquad \text{equation 4}$$

In the Monte Carlo algorithm, we separate out the total elastic scattering cross section from the angular distribution profile. Therefore, we only need to generate random angles that follow the same probability distribution as the true differential scattering cross section function. Thus, the most important aspects to be captured in our fit is that the width of the distribution (and how the width changes with energy), is accurately modeled. Here we compare the differential elastic scattering cross section from the NIST database [4] with that calculated by the Rutherford formula. To obtain a better fit, we scaled the factor $\beta$ by 0.9.

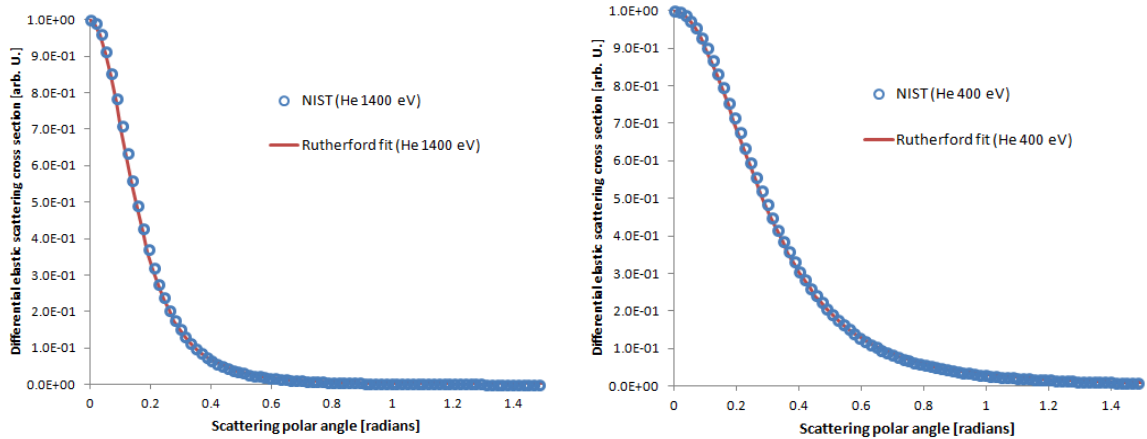

*Figure 6 - Differential elastic scattering cross section versus polar angle, for (left) 1400 eV and (right) 400 eV electrons scattered by He, as taken from the NIST database (blue circles) and fit using the Rutherford equation (red line).*

Here we see that the Rutherford fit does a good job matching the NIST data for Helium, which was calculated using more elaborate calculations. Note however, that the Rutherford line shape will not be able to reproduce the high-angle oscillations that are observed for heavier elements.

The differential scattering cross section shows how the cross section per unit solid angle varies as a function of polar angle. In the Monte Carlo algorithm we want to generate one random number for the azimuthal angle, and one random number for the polar angle. While the azimuthal angle can be a simple uniform random number, the polar angle must be uniform in spherical coordinates. Thus, in order to use a uniform random number to generate a random number in polar coordinates, we must perform a transformation. This process is described in the next section.

## 2.3 Random number generators

In the Monte Carlo algorithm, several random variables need to be generated. We use two methods for generating random numbers from a desired probability distribution function: 1) inversion sampling and 2) Metropolis algorithm. The method of inversion sampling is preferred because it is very efficient. However, inversion sampling requires that one has an invertible function for the cumulative probability distribution. In the case of inversion sampling, we make use of a uniform random number generator. In order to map the random number drawn from a uniform distribution onto a distribution in 3-dimensional space or using a different coordinate system, we need to write a transformation for each of the distributions we utilize. This procedure is described in detail in section 5.2-5.4. The trivial cases, where a transformation is not necessary are the random azimuthal angle, $\varphi$, which is chosen from a uniform distribution between 0 and $2\pi$. Furthermore, all randomly generated positions, in Cartesian coordinates are chosen from simple uniform distributions.

The distributions used here that required transformation are:

1) The exponential distribution (see section 5.3)

2) A distribution that is uniform over the surface of a sphere (see section 5.4)

3) The Rutherford distribution functions in spherical coordinates (see section 5.5)

4) A Lambertian distribution in spherical coordinates

In addition, a random number generator is needed for selecting random energy-loss amounts by choosing from the energy-loss function (see section 5.6). The derivations of the random number generators are provided in the additional information section.

## 3 Results

### 3.1 Validation of model

In order to verify that the Monte Carlo model is functioning correctly, we have run a few test cases to compare the results with expected results. First, we verify the results expected from the basic Poisson equation (equation 1). To do this, we constrain the electron paths, such that they all follow the shortest path from the sample to the nozzle (i.e. all electrons travel perpendicular to the sample's surface), they do not exhibit change in direction upon scattering, and they do not change their scattering cross section upon energy loss. Figure 7 shows the electron count versus number of times inelastically scattered, 1) resulting from a constrained Monte Carlo simulation and 2) calculated from the Poisson distribution. We see that the scattering statistics captured by the Monte Carlo model precisely match the result of the Poisson equation.

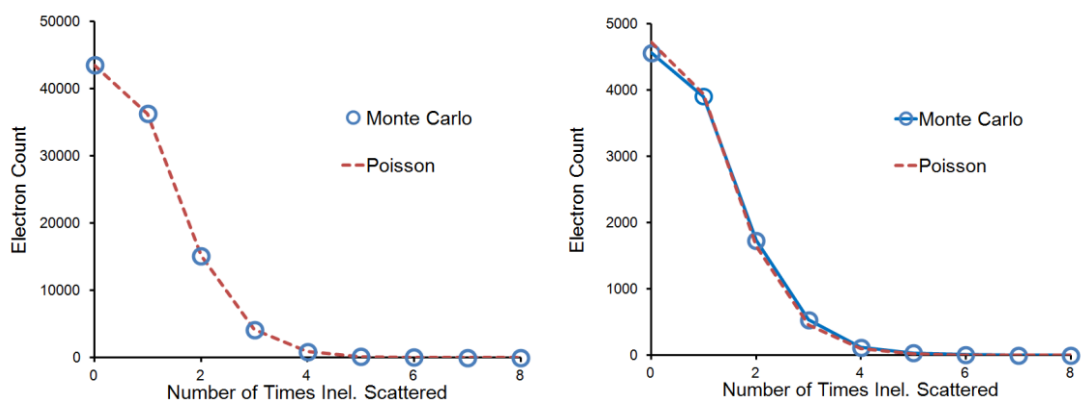

Figure 7 - (left) Plot of electron count versus number of times inelastically scattered as predicted using the Poisson equation (dashed-red line) and predicted using Monte Carlo simulations, where electrons travel perpendicular to the sample surface, with no angular deflection or change in cross section upon scattering. (right) Same comparison as with the left panel, except in the Monte Carlo calculation, deflection is allowed, the initial electron distribution is Lambertian, and cross section changes with kinetic energy.

If we switch on all of the physically realistic parameters, we see that there is a slight deviation from the Poisson calculations. The figure on the right shows a comparison between the unconstrained Monte Carlo calculation and the Poisson distribution. In these examples, the sample-nozzle distance was 0.3 mm, nozzle diameter was 0.3 mm, acceptance angle was 22°, the electrons were emitted from the surface over a 0.3 mm diameter spot, with a Lambertian angular emission profile. The inelastic and elastic scattering cross sections were allowed to change as a function of electron kinetic energy, and collisions were accompanied by angular deflection using a Rutherford distribution.

The deviation between the Monte Carlo calculation and the Poisson distribution can be corrected for, by replacing the sample-nozzle distance in the Poisson equation with the average path length. The sample-nozzle distance was 0.3 mm, and the average path length was 0.31, corresponding to a difference of ca. 3.5%.

The next test to verify is that the kinetic energy distributions of electrons that have been inelastically scattered *n* times, have the form of the primary electron distribution, convolved with the energy-loss function *n*-times. Here, we used a test initial kinetic energy distribution (resembling Ag3d), as shown in Figure 8-A. We ran a Monte Carlo simulation, where electrons with initial kinetic energies chosen from this distribution, were scattered through 0.3 mm of 25 mbar He, with an energy-loss function shown in Figure 8-B. The simulation was run for 1 million electrons. From the resulting distribution, we made selections of electrons that have been scattered *n* times, and plotted a histogram of their counts, versus kinetic energy, as shown in panel in Figure 8-C. For comparison, we calculated the line shapes of convolving the same primary energy distribution with the energy-loss function *n*-times. These line shapes are also plotted in Figure 8-C. Note, the intensities of the distributions from the Monte Carlo simulation were re-scaled so that the maximum height of the distribution matched with that of the convolved line-shapes. This was done for better visual comparison.

We can see that the electron distributions after inelastic scattering, resulting from the Monte Carlo simulation, match very well with the line shapes from convolution, indicating that the Monte Carlo algorithm is returning the expected behavior.

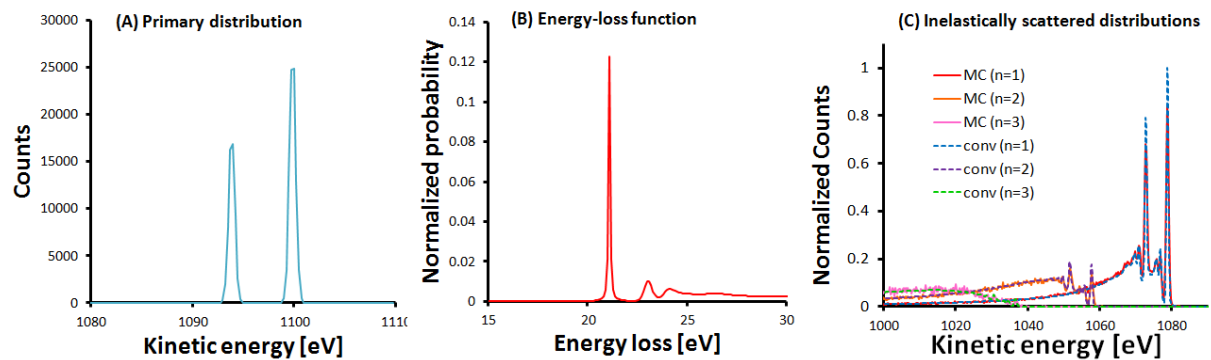

*Figure 8 - (left) initial electron kinetic energy distribution, (middle) energy-loss function, (right) line shapes resulting from a Monte Carlo simulation (solid lines) where the electrons with the initial kinetic energy distribution in the left panel were scattered through a scattering medium with the loss function shown in the middle panel, compared with multiple convolutions of the primary line shape with the energy-loss function line shape (dashed line). Note, the Monte Carlo line shapes were re-scaled to match the maximum height of the convolved line shapes.*

For additional validation tests, we ran simulations of electron scattering in the solid state. Thus, the electron source and scattering medium were the same object. There is an interesting result when applying the Poisson equation to solid-state scattering, which we can verify with the Monte Carlo simulation. From the Poisson equation, one can write the probability that an electron is scattered *n* times, when passing through a scattering medium of thickness *t*, where the scattering medium has cross section $\sigma$ and density $\rho$.

$$p_n = \left(\frac{1}{n!}\right)(t\rho\sigma)^n e^{-t\rho\sigma}$$
equation 5

In the context of solid-state scattering, we can think of *t* as roughly representing depth below the surface (strictly speaking it is path length), and $p_n(t)$ as the probability an electron from a depth *t*, has been scattered *n* times when it leaves the solid's surface. In photoemission measurements, electrons are generated at many depths below the surface, everywhere from several micrometers to nanometers below the surface. Thus, for each depth, we have a Poisson distribution, as shown in Figure 9. Note, this result ignores change in direction upon scattering.

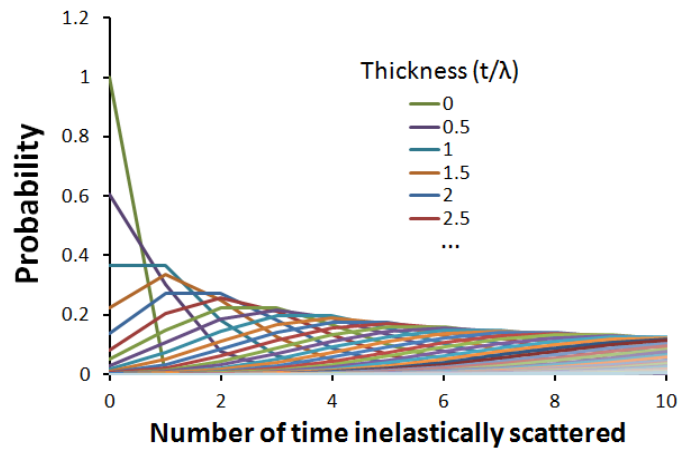

So in order to know the total number of electrons leaving the surface after *n* scattering events, we need to solve the following integral.

*Figure 9 - Plot of probability versus number of times inelastically scattered, for scattering medium thicknesses from t/λ = 0 to 20, as calculated from the Poisson equation (equation 1).*

$$I_n = I_0 \int_0^\infty \left(\frac{1}{n!}\right)(t\rho\sigma)^n e^{-t\rho\sigma}\, dt \qquad \text{equation 6}$$

Where $I_0$ represents the area density of electrons generated in the solid. If we solve this integral numerically we find that the integral is independant of *n*, and equates in all cases to $I_0/\rho\sigma$. What this means is that the number of electrons being detected is the same for all *n* (i.e. the intensity of the *n* = 0 signal, is the same as the intensity of the *n* = 1 signal, and the same as the *n* = 2 signal, etc.). This result is shown by the green line in the Figure 10. One can see that the electron count for all *n* is the same. In fact, this finding is only true if we assume no change in direction on scattering, and no change in inelastic scattering cross section with changes in kinetic energy. To test if the Monte Carlo algorithm returns the same result, we have simulated electron scattering in solid-state Ag in two scenarios: Where angular deflection and changing cross-section with kinetic energy 1) is turned on, and 2) is turned off.

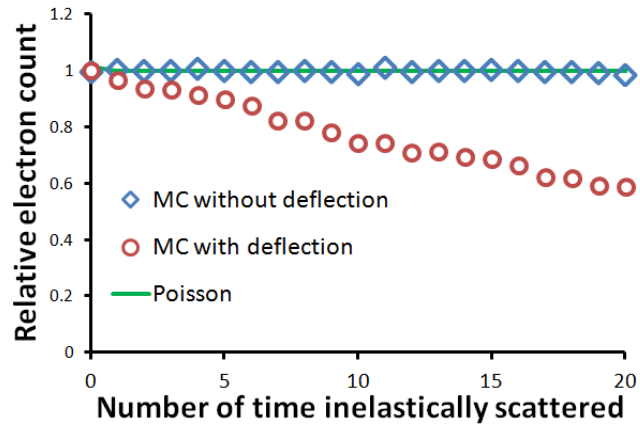

*Figure 10 - Plot of relative electron count versus number of times inelastically scattered for electrons scattering though bulk Ag, as determined from (blue diamond) Monte Carlo simulations where angular deflection and changing scattering cross section are disabled, (green line) Poisson statistics, (red circles) Monte Carlo calculations where angular deflection and changing scattering cross section are active.*

The result where deflection is turned off is shown as the blue diamonds in the figure. We see that the result derived from the Poisson equation is reproduced. In contrast, if we run the Monte Carlo simulation with deflection turned on, the the number of electrons escaping the surface decreases as a function of $n$.

Interestingly, if one were to ignore deflection (i.e. using only the intensities derived from Poisson behavior), then each of the $n$-convolutions, used to construct the total energy distribution, would have equal intensity, and the resulting line shape would yield a background signal that remains constant with decreasing kinetic energy, as shown in Figure 11. The blue curve is a test primary energy distribution. The red curve is the sum of convolutions between the primary distribution with an arbitary loss function n times, where n goes from 0 to 100. Each of the convolved line shapes has the same area (i.e. the result of the Poisson equation). One can see that the background converges to a flat signal at decreasing kinetic energies.

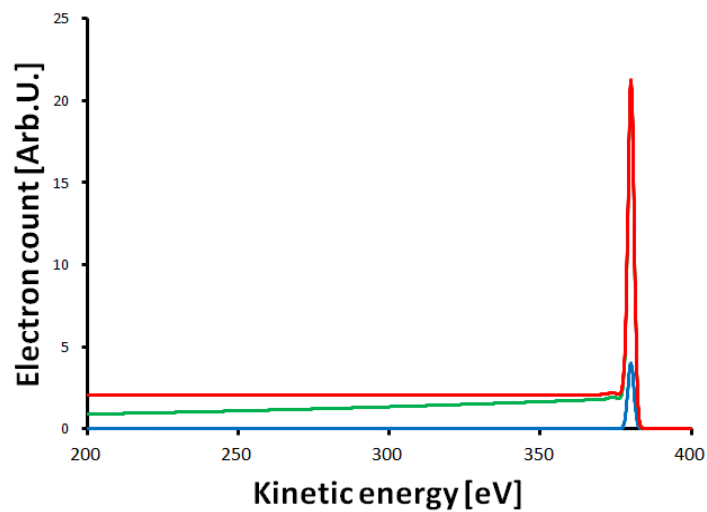

*Figure 11 - Spectra simulated using the Poisson equation, where a test initial kinetic energy distribution (blue) was convolved multiple times with a test energy loss function. (Red) The resulting sum of the convolved line shapes. (Green) The resulting sum of multiple convolutions, where the area of the $n^{th}$ convolved line shape is scaled by 0.9.*

If we re-scale the areas of the convolved line shapes, such that the areas decrease as a function of *n* ( in this case, $I_{n+1} = 0.9I_n$) then we produce a line shape that resembles the typical shape seen in XPS measurements. A comparison of simulated spectra using the Monte Carlo algorithm, where deflection is turned off, and where deflection is turned on, shows a similar trend (Figure 12). These results suggest that the effects of deflection during scattering, as well as changed in cross section with decreasing kinetic energy, result in fewer electrons reaching the surface than would be expected from Poisson statistics.

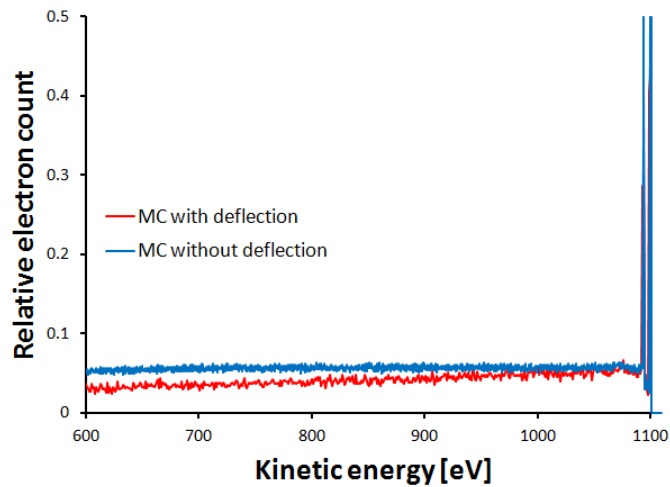

*Figure 12 - Monte Carlo simulations of background signals from Ag3d XPS spectra, where the background was simulated with (red) and without (blue) angular deflection and changing cross section during collision events.*

## 3.2 The effects of path-length distributions on using the Poisson distribution in gas-phase scattering

Angular deflection in scattering through the gas phase has an insignificant effect on the electron path length distribution (at least in any reasonable cases where there is still enough photoemission intensity reaching the spectrometer for a usable spectrum). Typically, for a real NAP-XPS measurement, one wants to minimize the number of times an electron is inelastically scattered. If more than a small fraction of the total intensity is inelastically scattered, the spectrum becomes unusable.

The main factor that influences the path-length distributions in NAP-XPS measurements–and gives rise to deviations from a simple Poisson distribution–is the geometry of the sample and nozzle. We have found that using the average path length in the Poisson equation provides a very good correction factor. Thus, we aim to use Monte Carlo simulations to find a trend between average path length, and relevant experimental configuration parameters, such as sample-nozzle distance, nozzle diameter, X-ray spot size, and nozzle acceptance angle. There are 5 variables that influence the path length distributions: sample-nozzle distance, pressure, acceptance angle, nozzle diameter, X-ray spot diameter. We can express average path length as a fraction of sample-nozzle distance, and call it normalized average path length. Then the normalized average path length is a function of 5 variables.

We used Monte Carlo simulations to determine the average path length for 2000 combinations of input parameters, and have fit the results to a 3rd degree polynomial. The resulting model results in 55 terms. The factors for each of the terms are shown below.

Table 1 - Values of pre-factors for polynomial fit $y = A_0 + A_1x_1 + A_2x_2 + ... + A_nx_1^2 + ... + A_nx_1^3$, where y: average path length / sample-nozzle distance, $x_1$: acceptance angle in degrees, $x_2$: nozzle diameter in mm, $x_3$: pressure in mbar, $x_4$: X-ray spot diameter in mm, $x_5$: sample-nozzle distance in mm.

| | | | | | |
|---|---|---|---|---|---|
| $A_0$ | 1.00177 | $x_5^2$ | -0.00799 | $x_2x_3^2$ | -5.62E-07 |
| $x_1$ | -0.00108 | $x_1^3$ | -5.59E-07 | $x_2x_3x_4$ | -7.13E-05 |
| $x_2$ | 0.071916 | $x_1^2x_2$ | 4.24E-05 | $x_2x_3x_5$ | 8.40E-05 |
| $x_3$ | 5.24E-05 | $x_1^2x_3$ | 9.34E-09 | $x_2x_4^2$ | 0.04856 |
| $x_4$ | 0.004496 | $x_1^2x_4$ | 1.30E-05 | $x_2x_4x_5$ | 0.049143 |
| $x_5$ | -0.0129 | $x_1^2x_5$ | -0.00012 | $x_2x_5^2$ | -0.25961 |
| $x_1^2$ | 9.34E-05 | $x_1x_2^2$ | -0.00115 | $x_3^3$ | 3.13E-08 |
| $x_1x_2$ | 0.004486 | $x_1x_2x_3$ | -1.57E-06 | $x_3^2x_4$ | 2.17E-06 |
| $x_1x_3$ | -6.12E-07 | $x_1x_2x_4$ | -0.0029 | $x_3^2x_5$ | 3.36E-06 |
| $x_1x_4$ | 0.001522 | $x_1x_2x_5$ | -0.00388 | $x_3x_4^2$ | 0.000166 |
| $x_1x_5$ | -0.00407 | $x_1x_3^2$ | -5.16E-09 | $x_3x_4x_5$ | -0.00034 |
| $x_2^2$ | -0.1987 | $x_1x_3x_4$ | 1.74E-06 | $x_3x_5^2$ | -7.59E-05 |
| $x_2x_3$ | 0.00015 | $x_1x_3x_5$ | 4.46E-07 | $x_4^3$ | -0.03198 |
| $x_2x_4$ | -0.08279 | $x_1x_4^2$ | 0.000118 | $x_4^2x_5$ | 0.069229 |
| $x_2x_5$ | 0.047427 | $x_1x_4x_5$ | 3.60E-05 | $x_4x_5^2$ | -0.17377 |
| $x_3^2$ | -3.72E-06 | $x_1x_5^2$ | 0.008349 | $x_5^3$ | 0.036797 |
| $x_3x_4$ | -8.54E-05 | $x_2^3$ | 0.046974 | | |
| $x_3x_5$ | 4.59E-05 | $x_2^2x_3$ | -2.25E-05 | | |
| $x_4^2$ | -0.01381 | $x_2^2x_4$ | 0.043375 | | |
| $x_4x_5$ | 0.095634 | $x_2^2x_5$ | 0.281051 | | |

Clearly the fit does not provide a very convenient equation. However, if one sets a constraint that sample-nozzle distance, nozzle diameter, and X-ray spot size are equal, then the average path length depends only on acceptance angle, and can be fit with a sigmoid, as shown in Figure 13. The equation for the fit is $L_{avg} \sim 1 + 0.077/(1 + \exp{(0.168(24.3 - \theta))})$, where $L_{avg}$ is the normalized average path length (i.e. average path length divided by sample-nozzle distance) and θ is the acceptance angle.

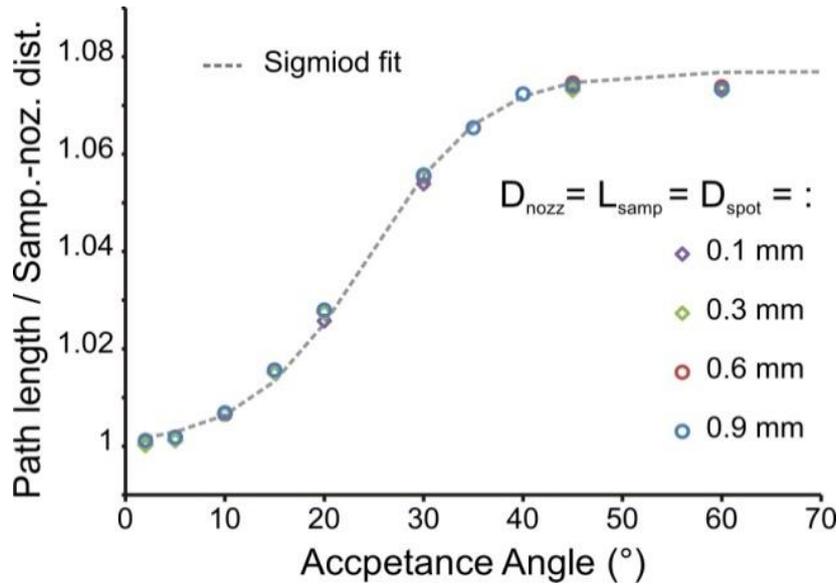

*Figure 13 - Plot of path length (t)/ sample-nozzle distance ($D_{samp}$) as a function of acceptance angle, for cases where $D_{nozz}$ = $L_{samp}$ = $D_{spot}$ = 0.1, 0.3, 0.6 and 0.9 mm. The results are from Monte Carlo simulations for 2 million electrons in 10 mbar He.*

With this curve, we have an accurate approximation of how much larger the average path length is than the sample nozzle distance. While this curve was fit for situations where $D_{nozz}$ = $L_{samp}$ = $D_{spot}$, deviations from this behavior result in only minimal deviations from this curve. Acceptance angle has the strongest effect on average path. For example, with $D_{nozz}$ = $L_{samp}$ = $D_{spot}$ = 0.3 mm, using a 22° acceptance angle, the average path length is 3% larger than the sample-nozzle distance. If one then compares with a situation where $D_{nozz}$ = 0.3 mm, $L_{samp}$ = 1 mm, $D_{spot}$ = 1 mm then the average path length is 2% larger than the sample-nozzle distance.

## 4. Conclusion

We have used Monte Carlo simulations of inelastic electron scattering through a gas phase to test the accuracy of using the Poisson equation for determining scattering statistics. We find that, due to sample and spectrometer geometrical configurations, electrons exhibit a distribution of path lengths as they travel through the gas. By replacing the distance term in the Poisson equation with the average path length, one can effectively correct for this error. However, the average path length is a function of 5 parameters, making it not practical to routinely calculate. We found that, after normalizing the average path length by dividing it by sample-nozzle distance, the most influential factor is acceptance angle. We fit the behavior of average path length vs. acceptance angle to provide a practical equation for estimating the correction factor to be used in the Poisson equation.

## 5 Additional info
### 5.1 Line tracing and shape intersections

In the python script used for this algorithm there are multiple kinds of shapes. These shapes are used for several purposes: 1) as the source of electrons, 2) as the detector, 3) as the scattering medium. A sphere or calendar is generally used as the boundary. A disc is used as the detector, and a disc is generally used as the source. The scattering medium could be a disc or a sphere. In the case where we are simulating solid state scattering, we would use a disc, where the source is inside the disc. In the case of gas phase scattering, the scattering medium would be a sphere inside of the detector sphere. The shapes have common methods, for example, to detect if an arbitrary point is inside the shape or outside the shape. They also have methods to detect the point of intersection between a line and the shape.

For the sphere, the equations for finding an intersecting point are as follows:

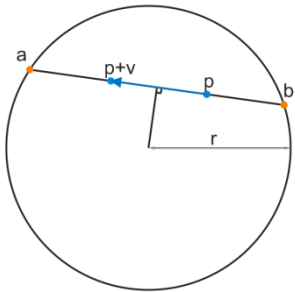

If we have a sphere of radius R, a point inside the sphere $p_0$ = [$p_x$, $p_y$, $p_z$] and a vector defining the velocity of a particle $v$ = [$v_x$, $v_y$, $v_z$]. The line is defined by $l$ = $v$t + $p_0$ and we want to find the Cartesian coordinates of the point A, where the line intersects with the sphere.

The Cartesian coordinates of the line are given by:

$$x = p_x + tv_x$$
$$y = p_y + tv_y$$
$$z = p_z + tv_z$$

## 5.2 Probability distributions under transformation

The derivations for probability distributions under transformation follow from reference [12].
In general, consider two sets, A and B. $p_A$ is a probability density on A, and $f$ is a function that transforms A to B. If one samples $x$ from A according to the probability density $p_A$, we wish to know the form of the probability density $f(x)$.

The function $f$ sends points on the interval [$x$, $x+\delta x$] to [$f(x)$, $f(x+\delta x)$]. By Taylor series expansion, we can write $f(x+\delta x) = f(x) + f'(x)\delta x$

We can write the mass probability on in interval [$x$, $x+\delta x$] on A as $p_A(\underline{x})\delta x$

Then the probability density on B will be this mass divided by the differential interval on B.

$p_B(f(x)) = p_A(x)\delta x/|f'(x)|dx = p_A(x)/|f'(x)|$

If we then choose our density distribution on A to be uniform on the interval [0,1), we can write

$p_B(f(x)) = 1/|f'(x)|$

$p_B(f(x))|f'(x)| = 1$

In the inversion method, we require the cumulative distribution function ($P_B$) of our target distribution.

$$P_B(y) = \int_{-\infty}^{y} p_B(t) dt$$

Now we have:

$$\left(P_B(f(x))\right)' = p_B(f(x))|f'(x)|$$

and

$$\left(P_B(f(x))\right)' = 1$$

If we integrate both sides, and assume $P_B(f(0)) = 0$, we have

$$\int_0^x \left(P_B(f(t))\right)' dt = \int_0^x 1 dt$$
$$P_B(f(x)) = x$$
$$f(x) = P_B^{-1}(x)$$

With this equation, we can pick a random number between 0 and 1 on a uniform distribution, and get back a point on our target distribution.

### 5.3 The exponential distribution

In the MC algorithm, when selecting the distance an electron travels to the next collision, we use Beer's law, which states:

$p(x) = \lambda e^{-\lambda x}$

where lambda is the mean free path, and can be calculated from the total cross section and density as $1/(\sigma_{tot} * \rho)$, where $\sigma_{tot}$ is the total scattering cross section and $\rho$ is the scatterer density.

For the exponential distribution, the cumulative distribution function is:

$$P(x) = 1 - e^{-\lambda x}$$

Then the distribution can be sampled as: $x = \frac{-\ln(1-y)}{\lambda}$

Where y is a uniform random number between 0 and 1.

### 5.4 Uniform distribution on a sphere

For initializing the electron velocity, the Monte Carlo algorithm can use a random direction that has equal probability in all directions. We choose the azimuthal angle uniformly between 0 and 2π, however, the polar angle distribution density should be lower at the poles than at the equator. Thus, one cannot choose the polar angle from a uniform distribution.

For this case, we need to transform a distribution on **R**² to **R**³. In general:

$$f(u, v) = \begin{bmatrix} f_x(u, v) \\ f_y(u, v) \\ z(u, v) \end{bmatrix}$$

The partial derivatives are:

$$f_u = \begin{bmatrix} \frac{\partial f_y}{\partial u}(u, v) \\ \frac{\partial f_y}{\partial u}(u,v) \\ \frac{\partial f_z}{\partial u}(u,v) \end{bmatrix}, f_v = \begin{bmatrix} \frac{\partial f_y}{\partial v}(u, v) \\ \frac{\partial f_y}{\partial v}(u,v) \\ \frac{\partial f_z}{\partial v}(u,v) \end{bmatrix}$$

The probability distribution transformation becomes:

$$p_B(f(u, v)) = \frac{p_A(u, v)}{\sqrt{EG - F^2}}$$

where we define:

$$E = f_u \cdot f_u$$

$$F = f_u \cdot f_v$$

$$G = f_v \cdot f_v$$

For the case of spherical coordinates, our function that transforms polar coordinates to Cartesian coordinates is:

$$f(\varphi, \theta) = \begin{bmatrix} f_x(\varphi, \theta) \\ f_y(\varphi, \theta) \\ z(\varphi, \theta) \end{bmatrix} = \begin{bmatrix} \sin \theta \cos \varphi \\ \sin \theta \sin \varphi \\ \cos \theta \end{bmatrix}$$

and the partial derivatives are:

$$f_\theta = \begin{bmatrix} \cos \theta \cos \varphi \\ \cos \theta \sin \varphi \\ -\sin \theta \end{bmatrix}, f_\varphi = \begin{bmatrix} -\sin \theta \sin \varphi \\ \sin \theta \cos \varphi \\ 0 \end{bmatrix},$$

We have then:

$$E = 1, F = 0, G = \sin^2 \theta$$

and

$$\sqrt{EG - F^2} = |\sin \theta|$$

And our probability distribution in Cartesian coordinates is:

$$p_B(f(\theta, \varphi)) = \frac{p_A(\theta, \varphi)}{\sin \theta}$$

Then, to uniformly sample a sphere, we want the probability density to be uniform over the entire area of the sphere. The area of a sphere in steradians is 4π. Thus, the density of points on the sphere should be 1/4π, and:

$$\frac{1}{4\pi} = \frac{p_A(\theta, \varphi)}{\sin \theta}$$

$$p_A(\theta, \varphi) = \frac{\sin \theta}{4\pi}$$

Since we will sample the azimuthal angle separately from the polar angle, and the azimuthal angle is sampled uniformly from 0 to 2π, we can write:

$$\frac{p_A(\theta)}{2\pi} = \frac{\sin \theta}{4\pi}$$

$$p_A(\theta) = \frac{\sin \theta}{2}$$

Then the cumulative distribution function is:

$$\int_0^\theta p_A(\alpha) d\alpha = \frac{1 - \cos \theta}{2}$$

and the inverse of the cumulative distribution function is:

$$P_\theta^{-1} = \cos^{-1}(1 - 2\xi)$$

Thus we select θ:= cos⁻¹(1-2ξ), where ξ is a uniform random number between 0 and 1.

### 5.5 Rutherford distribution in spherical coordinates

The result of this derivation is used frequently in Monte Carlo simulations in the published literature, however, a derivation of it is difficult to find. The Rutherford distribution was used to choose polar angles in each MC step. The Rutherford distribution represents the cross section per unit solid angle that an electron is scattered into a given polar angle.

$$\frac{d\sigma}{d\Omega} = \frac{Z^2 e^4}{4E^2(1 - \cos \theta + 2\beta_N)^2}$$

Substituting the Rutherford distribution into our target probability density function, we get:

$$\frac{d\sigma}{d\Omega} \frac{1}{C} = \frac{p_A(\theta, \varphi)}{\sin \theta}$$

$$\frac{Z^2 e^4}{4E^2} \frac{\sin \theta}{4\left(\sin^2 \frac{\theta}{2} + \beta_N\right)^2} \frac{1}{2\pi C} = p_A(\theta)$$

The factor C is there to ensure that the total probability sums to one. It is obtained by computing the integral for the target distribution over its entire range of validity (0,π).

$$C = \int_0^\pi \frac{d\sigma}{d\Omega} \sin\theta \, d\theta$$

Then our cumulative distribution function becomes...

$$P_B(\theta) = \frac{\int_0^\theta \frac{d\sigma}{d\Omega} \sin\theta' \, d\theta'}{\int_0^\pi \frac{d\sigma}{d\Omega} \sin\theta' \, d\theta'}$$

First we plug our expression for the differential cross section into the integral...

$$P_B(\theta) = \frac{\int_0^\theta \frac{\sin\theta'}{\left(\sin^2\frac{\theta'}{2} + \beta_N\right)^2} d\theta'}{\int_0^\pi \frac{\sin\theta'}{\left(\sin^2\frac{\theta'}{2} + \beta_N\right)^2} d\theta'}$$

Then we evaluate the definite integrals...

$$P_B(\theta) = \frac{\left.\frac{4}{\cos\theta' - 2\beta - 1}\right|_0^\theta}{\left.\frac{4}{\cos\theta' - 2\beta - 1}\right|_0^\pi}$$

$$P_B(\theta) = \frac{\frac{4}{\cos\theta - 2\beta - 1} + \frac{2}{\beta}}{\frac{4}{-2 - 2\beta} + \frac{2}{\beta}}$$

We simplify the denominator...

$$P_B(\theta) = \frac{\frac{4}{\cos\theta - 2\beta - 1} + \frac{2}{\beta}}{\frac{-2}{1+\beta} + \frac{2}{\beta}}$$

$$P_B(\theta) = \frac{\frac{4}{\cos\theta - 2\beta - 1} + \frac{2}{\beta}}{2\left(\frac{-1}{1+\beta} + \frac{1}{\beta}\right)}$$

$$P_B(\theta) = \frac{\frac{4}{\cos\theta - 2\beta - 1} + \frac{2}{\beta}}{2\left(\frac{-\beta}{(1+\beta)\beta} + \frac{1+\beta}{(1+\beta)\beta}\right)}$$

$$P_B(\theta) = \frac{\frac{4}{\cos\theta - 2\beta - 1} + \frac{2}{\beta}}{2\left(\frac{1}{(1+\beta)\beta}\right)}$$

$$P_B(\theta) = \frac{\frac{2}{\cos\theta - 2\beta - 1} + \frac{1}{\beta}}{\left(\frac{1}{(1+\beta)\beta}\right)}$$

Now we simplify the numerator...

$$P_B(\theta) = \frac{\frac{2\beta + \cos\theta - 2\beta - 1}{(\cos\theta - 2\beta - 1)\beta}}{\left(\frac{1}{(1+\beta)\beta}\right)}$$

$$P_B(\theta) = \frac{\frac{\cos\theta - 1}{(\cos\theta - 2\beta - 1)\beta}}{\left(\frac{1}{(1+\beta)\beta}\right)}$$

$$P_B(\theta) = \frac{(\cos\theta - 1)\bigl((1+\beta)\beta\bigr)}{(\cos\theta - 2\beta - 1)\beta}$$

$$P_B(\theta) = \frac{(\cos\theta - 1)(1+\beta)}{(\cos\theta - 1 - 2\beta)}$$

Now try to invert the equation to get an expression for theta as a function of R, where R is a random number between 0 and 1

$$R = \frac{(\cos\theta - 1)(1+\beta)}{(\cos\theta - 1 - 2\beta)}$$

If we plot this relationship, we see that it should be invertible

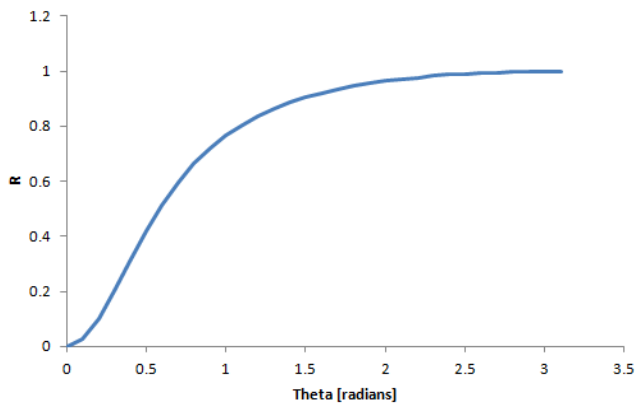

$$R(\cos\theta - 1 - 2\beta) = (\cos\theta - 1)(1+\beta)$$

$$R\cos\theta - R - 2\beta R = \cos\theta + \beta\cos\theta - 1 - \beta$$

$$-\cos\theta - \beta\cos\theta + R\cos\theta = -1 - \beta + R + 2\beta R$$

$$\cos\theta\,(-1 - \beta + R) = -1 - \beta + R + 2\beta R$$

$$\cos\theta = \frac{-1 - \beta + R + 2\beta R}{-1 - \beta + R}$$

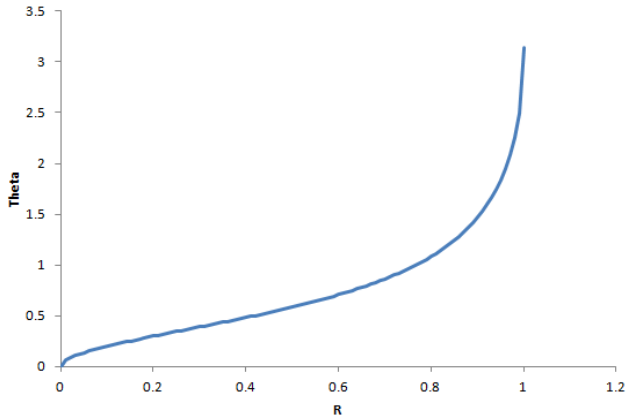

Therefore, we can randomly obtain an angle, θ, that follows our desired distribution, by choosing a random number R from the uniform distribution on the interval [0,1] as…

$$\theta = \cos^{-1}\frac{R(1 + 2\beta) - 1 - \beta}{R - 1 - \beta}$$

If we assume the sphere is centered at (0,0,0) then the equation of the sphere is
$$x^2 + y^2 + z^2 = R^2$$

If we substitute the coordinates of the line into the equation of the sphere, we arrive at:
$$at^2 + bt + c = 0$$
where
$$a = v_x^2 + v_y^2 + v_z^2$$
$$b = 2(v_x p_x + v_y p_y + v_z p_z)$$
$$c = p_x^2 + p_y^2 + p_z^2 - 2(p_x + p_y + p_z) - R^2$$
The solutions are $t = (-b +/- \sqrt{b^2 - 4ac})/2a$

## 5.6 The Metropolis algorithm for drawing from the energy-loss function

The Metropolis algorithm, for randomly drawing from a probability distribution function, is an efficient means of generating random number when there is no invertible analytical function for the probability density function. We use it here to randomly select the amount of energy lost during electron-molecule collision, by drawing from the molecule's energy-loss function.

In the metropolis algorithm, one has some probability distribution function, $f(x)$, that is, given some value of x, $f(x)$ tells the probability of the event x. In the Metropolis algorithm, one randomly picks a value of x (within some bounds), and sets it to the variable *current_x*. Then one calculates the corresponding value of $f(current\_x)$ and sets it to the variable *current_y*. One then selects a new trial value of x, called *trial_x*. This is done by taking a random step away from *current_x*, such that *trial_x* = *rand* * *step* + *current_x*, where rand is a uniform random number between 1 and -1, and step is a constant. One then calculates *trial_y* = $f(trial\_x)$. One then compares the values of *trial_y* and

*current_y* by computing the ratio *R* = *trial_y* / *trial_y*. One then draws another rand number, *r*, from a uniform distribution between 0 and 1, and compares the values of *R* and *r*. If *r* <= *R*, then the move is accepted, and *current_x* takes on the value of *trial_x*, and the value of *trial_x* is appended to a list of accepted *x* values. The procedure is repeated *n* times, and a histogram of the list of accepted x values should have approach the form of the real probability distribution.

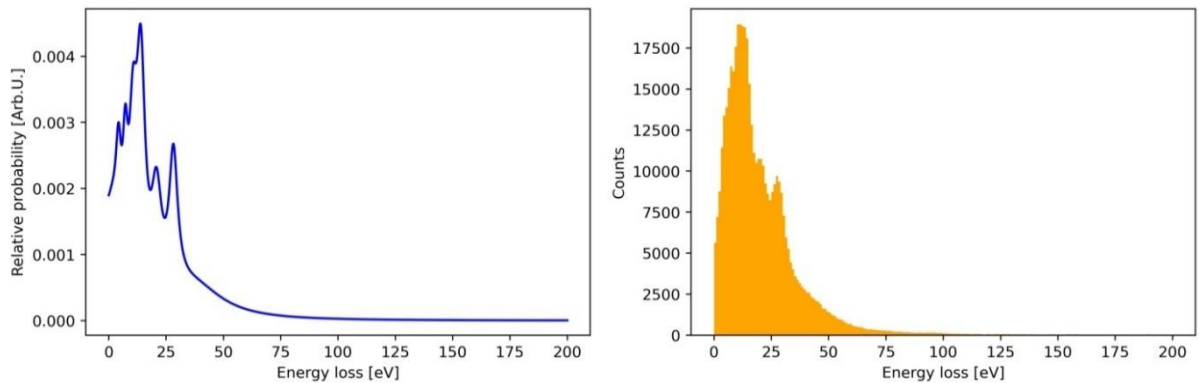

The figure above is a comparison of the real probability density function (left) with a histogram of randomly chosen values using the Metropolis algorithm.

An alternative method for sampling, similar to inversion sampling, but suitable for distributions for which there is no analytical function is to plot the cumulative distribution function, and normalize it, so that its maximum value is 1, and minimum is 0. Then store the energy loss values in one array and the cumulative probabilities in another array. Pick a uniformly distributed random number between 0 and 1. Find the index of the value in the cumulative distribution array that is closest to the random number. Then take the corresponding energy loss value (i.e. the value in the energy loss array with the corresponding index). This method is substantially faster than the Metropolis algorithm for a 1D probability density function.